\documentclass[aps,prb,showpacs,preprint]{revtex4}

\usepackage{amsmath}% math extension
\usepackage{graphicx}% include figure files

\begin{document}

\title{Nonlinear hopping transport in ring systems and open channels}
\author{Mario Einax}
\affiliation{Institut f\"ur Physik,
Technische Universit\"at Ilmenau, 98684 Ilmenau,Germany}
\author{Martin K\"orner}
\affiliation{Institut f\"ur Physik,
Technische Universit\"at Ilmenau, 98684 Ilmenau,Germany}
\author{Philipp Maass}
\affiliation{Fachbereich Physik,
Universit\"at Osnabr\"uck, Barbarastrasse 7, 49069 Osnabr\"uck, Germany}
\email{philipp.maass@uni-osnabrueck.de}
\homepage{http://www.tu-ilmenau.de/theophys2}
\author{Abraham Nitzan}
\affiliation{School of Chemistry, Tel Aviv University, Tel Aviv 69978,
Israel}

\date{June 16, 2009}

\begin{abstract}

  We study the nonlinear hopping transport in one-dimensional rings
  and open channels. Analytical results are derived for the stationary
  current response to a constant bias without assuming any specific
  coupling to the external fields. It is shown that anomalous large
  effective jump lengths, as observed in recent experiments by taking
  the ratio of the third order nonlinear and the linear conductivity,
  can occur already in ordered systems. Rectification effects due to
  site energy disorder in ring systems are expected to become
  irrelevant for large system sizes. In open channels in contrast,
  rectification effects occur already for disorder in the jump
  barriers and do not vanish in the thermodynamic limit. Numerical
  solutions for a sinusoidal bias show that the ring system provides a
  good description for the transport behavior in the open channel for
  intermediate and high frequencies. For low frequencies temporal
  variations in the mean particle number have to be taken into account
  in the open channel, which cannot be captured in the more simple
  ring model.

\end{abstract}


%66.30.H- Self-diffusion and ionic conduction in nonmetals 
%05.60.Cd Classical transport
%66.10.Ed Ionic conduction 

\maketitle

\section{Introduction}

The particle transport in one-dimensional systems is of vital interest
for many problems in physics and biology. A prominent example is the
electron or hole transport in the operation of conducting nanowires,
including molecular wires.\cite{Nitzan/etal:2003} In such systems
transport can be dominated by quantum mechanical tunneling or band
motion (the coherent transport limit) but many systems belong to the
hopping transport limit, where conduction is a manifestation of
succession of many incoherent hopping
steps.\cite{Nitzan/etal:2001,Romano/etal:2007} For example, both
conduction mechanisms were observed in different DNA
sequences.\cite{Xu/etal:2004} One-dimensional hopping motion is also
the decisive transport mechanism in ion conduction through membrane
channels \cite{Hille:2001,Berneche/Roux:2003,Graf/etal:2004} and
unidirectional motion of motor proteins along
filaments.\cite{MacDonald/etal:1968,Frey/Kroy:2005} In the connection
of the latter example much attention have received recently boundary
driven phase transitions in one-dimensional lattice gases with site
exclusion and asymmetric hopping dynamics, commonly referred to as
``asymmetric site exclusion process'' (ASEP), or, in case of
unidirectional transport, as ``totally asymmetric site exclusion
process '' (TASEP) -- for reviews, see Refs.\ \onlinecite{Spohn:1981,
  Derrida/etal:1997, Schuetz:2001}. Recently, properly modified models
\cite{Zilman:2009} were applied to describe the transport of
single-stranded DNA segments through
nanochannels.\cite{Kohli/etal:2004}

The treatment of one-dimensional systems is moreover frequently used
as a starting point for describing transport processes in higher
dimensions, since it often allows one to derive analytical results. In
transferring essential results to higher dimensions one has, however,
to be careful. An example is the tracer diffusion in one-dimensional
hard-core lattice gases, which exhibits a subdiffusive behavior for
long times that originates from the fact that particles cannot pass
each other in one
dimension.\cite{Harris:1965,Beijeren/etal:1983,Kollmann:2003}

In this work we will study the thermally activated hopping conduction
in one-dimensional lattices for non-interacting particles in arbitrary
energy landscapes. In particular we consider the nonlinear transport
in strong static and periodic fields. For couplings $\propto\exp(\pm
u/2)$ of the external bias $u$ to the bare hopping rate, this problem
was first studied for ring systems (periodic boundary conditions) in
Ref.~\onlinecite{Kehr/etal:1997}. An exact result for the stationary
current was derived, in generalization of an analogous treatment for
Brownian dynamics.\cite{Ambegaokar/etal:1969} As a particularly
interesting feature, rectification effects were shown to be present
for energy landscapes with site energy disorder.

The problem got renewed interest recently for describing measurements
on thin glassy electrolytes under high voltages,\cite{Roling:2001,
  Murugavel/etal:2005, Heuer/etal:2005, Roling/etal:2008} which allow
one to reach the weak nonlinear regime, $u=qEa/k_{\rm B}T\simeq1$,
where $q$ is the charge of the mobile ions, $a$ is a typical hopping
distance of 2-3\AA, $E$ is the applied electric field, and $k_{\rm
  B}T$ the thermal energy. In these experiments no rectification were
observed so far, meaning that the current turned out to be an odd
function of the applied field. On the other hand, these measurements
can be used to determine an effective length scale $a_{\rm eff}$ when
analyzing the ratio $\sigma_3/\sigma_1$ of the third order nonlinear
conductivity $\sigma_3$ to the linear conductivity $\sigma_1$ (cf.\
Eq.~\ref{eq:aeff} below). This length $a_{\rm eff}$ appears to be
unphysically large if it is compared to typical jump lengths
$a\simeq2-3$~\AA. Such comparison is motivated by the result
\cite{Mott/Davis:1979} $j_{\rm dc}\propto\sinh(qEa/2k_{\rm B}T)$,
which applies to the most simple situation of single-particle hopping
in an ordered system with the aforementioned coupling $\propto\exp(\pm
u/2)$ of the bias to the bare hopping rates (see below). For different
glassy electrolytes $a_{\rm eff}$ either increases or decreases with
$T$ (in the temperature ranges studied a linear behavior was
observed). It was also found that $\sigma_3>0$, while $\sigma_5$ has
different sign for different glass compositions. In the
frequency-dependent response the real part $j_3'(\omega)$ of the third
order harmonics $\hat j_3(\omega)$ has a negative sign for low
frequency. With increasing frequency, $j_3'(\omega)$ increases and
becomes positive close to the onset frequency of the dispersive part
in the first order harmonics $j_1'(\omega)$ (which gives the linear
response conductivity $\sigma_1'(\omega)$).

Taking disorder averages \cite{Heuer/etal:2005} of the analytical
expression for the current derived in
Ref.~\onlinecite{Kehr/etal:1997}, it was suggested that the large
values of $a_{\rm eff}$ have their origin in the spatial variation of
hopping rates in the glassy material. Moreover, based on a small $u$
expansion, it was predicted that $a_{\rm eff}\propto N^{1/2}$, where
$N\simeq L/a$ is the number of sites of the film sample in field
direction. However, this result followed when expanding terms as
$\exp(Nu)$ in the analytical result for the current in power of $Nu$.
Since the nonlinear transport becomes relevant for $u\gtrsim1$, and
$N$ should be significantly larger than one (to avoid boundary
effects), this expansion in powers of $Nu$ is in general not
appropriate. Rather one should take the thermodynamic limit
$N\to\infty$ before carrying out the small $u$ expansion of the
current,\cite{Maass:2006} which can yield non-analyticities in the
current response. It was argued \cite{Roling/etal:2008} that these
non-analyticities could spoil the analysis of nonlinear conductivities
based on odd powers in the field amplitude, as they are commonly
employed in experiments.

An open question is whether the rectification effects occurring in
finite systems are present also in the thermodynamic limit.
Intuitively, one would expect that in the absence of long-range
correlations in the energy landscape (i.e.\ correlations decaying
faster than 1/distance), self-averaging effects suppress rectification
properties the more the larger the system size becomes. As a
consequence one would predict rectification effects to disappear in
the thermodynamic limit. While this is in agreement with experimental
observations (for sample thicknesses so far studied), it has not yet
been demonstrated by theoretical analysis. To avoid the problem of
possible rectification effects and to enforce that the current is an
odd function of $u$, energy landscapes with point symmetry were
considered in Refs.~\onlinecite{Heuer/etal:2005,Roling/etal:2008}.
However, the constraint of point symmetry implicitly introduces
long-range correlations in the energy landscape and it is questionable
if such procedure is suitable to describe real experimental
situations.

In this work we will treat the following open problems:
\begin{list}{}{\leftmargin1.5em}{\setlength{\itemsep}{-1cm}}
  {\setlength{\parsep}{-0.5cm}}

\item[(1)] Analytical results for the stationary current in ring
  system with $M$ sites were derived up to now for the coupling
  $\propto\exp(\pm u/2)$ of the external bias $u=qEa/k_{\rm B}T$ to
  the bare rates (rates in the absence of the external driving). This
  rate emerges naturally when approaching the hopping limit of the
  overdamped Brownian dynamics (Smoluchowski equation) of
  noninteracting particles. However, in interacting many-particle
  systems more complicated couplings of the rates to the external
  field can be imagined, when mapping the dynamics to an effective
  one-particle hopping process in a renormalized energy landscape. We
  therefore derive the stationary current for arbitrary couplings, and
  discuss in more detail the behavior for jump rates obeying the
  condition of detailed balance. We find that is then possible to
  obtain already in an ordered system effective lengths scales $a_{\rm
    eff}$ significantly larger than the jump length $a$. Hence it
  appears that not only the disorder affects $a_{\rm eff}$.

\item[(2)] As outlined above, for relating the theoretical results to
  experiments in the nonlinear regime, one should first perform the
  thermodynamic limit $M\to\infty$ before expanding the current in
  powers of the field amplitude. By performing this limit we also
  clarify the role of rectification effects for large $M$.

\item[(3)] For ring systems it is unclear how the periodic boundary
  conditions affect the stationary current. We therefore study the
  analogous problem in an open channel, where particles are injected
  and ejected from two particle reservoirs on the left and right side
  with electrochemical potentials $\mu_{\rm L}$ and $\mu_{\rm R}$,
  respectively. The rates for the local exchange of particles with the
  reservoirs fulfill detailed balance with respect to the
  grand-canonical ensembles associated with $\mu_{\rm L}$ and
  $\mu_{\rm R}$. We will treat the linear limit of the rate equations
  in this work to avoid boundary induced phase transitions as
  occurring in ASEPs or TASEPs. \cite{Spohn:1981, Derrida/etal:1997,
    Schuetz:2001}

\item[(4)] Up to now the time-dependent nonlinear current response has
  been rarely studied.\cite{comm:time-dependent} Here we will
  investigate for both the ring systems and open channels this
  time-dependent response to a sinusoidal driving with large field
  amplitude $E_0$ by numerically solving the corresponding rate
  equations for the occupation probabilities. The data are analyzed,
  by using standard Fourier analysis, in terms of harmonics $\hat
  j_n(\omega)$ of $n$th order. We present results for spatially
  uncorrelated barrier energies with uniform distributions and discuss
  the relation of the harmonics in the ring and open channel with
  respect to different frequency regimes.

\end{list}

\section{Transition Rates and Energetic Disorder}\label{sec:transrates}

For convenient notation, we define $k_{\rm B}T$ as the energy unit in
the following, $k_{\rm B}T=1$. In a disordered energy landscape with
site energies $\epsilon_k$ and energy barriers $U_{k,k+1}=U_{k+1,k}$
between sites $k$ and $k+1$, the rates $\Gamma_k^+(t)$ and
$\Gamma_{k+1}^-(t)$ are considered to be functions of $\epsilon_k$,
$\epsilon_{k+1}$, and $U_{k,k+1}$. In addition they depend on the
external bias $u(t)$, which we assume to be homogenous over the ring
or channel, corresponding to a linear decrease of the external
potential. If the rates obey detailed balance at each time instant,
their ratio $\eta_k(t)$ is given by
\begin{equation}
\eta_k(t)=\frac{\Gamma_k^+(t)}{\Gamma_{k+1}^-(t)}=
\exp(-\Delta E_{k,k+1})\,,
\label{eq:detailed-balance}
\end{equation}
where
\begin{equation}
\Delta E_{k,k+1}=\epsilon_{k+1}-\epsilon_k-u\,.
\label{eq:deltaE}
\end{equation}
In the presence of screening effects, the assumption of a constant
potential gradient is not valid, leading to a bias depending on $k$.
The analytical formulae derived in the following sections can be
generalized to this situation.

To illustrate our findings we will consider two types of rates and two
types of energetic disorder. For the rates these are the ``exponential rates''
\begin{equation}
\Gamma_k^+=\frac{\gamma}{2}\exp(-U_{k,k+1})\exp(-\Delta E_{k,k+1}/2)
\label{eq:exp-rates}
\end{equation}
and the Glauber rates \cite{Glauber:1963}
\begin{equation}
\Gamma_k^+=\frac{\gamma}{2}\exp(-U_{k,k+1})
\left[1+\tanh\left(\frac{\Delta E_{k,k+1}}{2}\right)\right]
=\gamma\,\frac{\exp(-U_{k,k+1})}{1+\exp(-\Delta E_{k,k+1})}\,,
\label{eq:glauber-rates}
\end{equation}
where $\gamma$ is a bare jump rate. For the energetic disorder, we
consider either pure barrier disorder (all $\epsilon_k=0$), or pure
site energy disorder (all barriers $U_{k,k+1}=0$). The barrier and
site energies are uncorrelated random variables drawn from box
distributions, $U_{k,k+1}\in[0,\Delta_U]$ and
$\epsilon_i\in[-\Delta_\epsilon/2,\Delta_\epsilon/2]$ with widths
$\Delta_U$ and $\Delta_\epsilon$, respectively.

\section{Conduction in ring systems}\label{sec:ring}

We study the nearest neighbor hopping of one particle on a ring with
$M$ sites $i=1,\ldots,M$. The rates for a jump from site $i$ backward
and forward at time $t$ are denoted as $\Gamma_i^-(t)$ and
$\Gamma_i^+(t)$, respectively. The probabilities $p_i(t)$ for the
particle to be on site $i$ at time $t$ obey the rate equations
\begin{equation}
\dot p_i=j_{i-1,i}(t)-j_{i,i+1}(t)\,,\hspace{2em} i=1,\ldots,M\,,
\label{eq:rate-eq-ring}
\end{equation}
with the local currents
\begin{equation}
j_{i,i+1}(t)=\Gamma_i^+(t)p_i(t)-\Gamma_{i+1}^-(t)p_{i+1}(t)\,,
\hspace{2em} i=1,\ldots,M\,. \label{eq:local-j-ring}
\end{equation}
In writing Eqs.~(\ref{eq:rate-eq-ring},\ref{eq:local-j-ring}) and
further equations below we implicitly assume that the periodic
boundary conditions are taken into if the index $i$ falls out of
the range $1,\ldots,M$, i.e.\ $p_{i+M}(t)=p_i(t)$,
$\Gamma_{i+M}^\pm(t)=\Gamma_i^\pm(t)$,
$j_{i+M,i+1+M}(t)=j_{i,i+1}(t)$, etc. The rate equations preserve the
normalization $\sum_{i=1}^Mp_i(t)=1$.

Due to the normalization of the occupation probabilities to one
particle, the current $j$ refers to the single particle current. If we
consider a fixed number density $n$ per lattice site of
non-interacting particles, the total current is
\begin{align}
\label{eq:total_current}
J&=nMj \, .
\end{align}
In the case of charged particles the corresponding charge current per
lattice site is $qJ$ and the charge current density $qJ/A$, where $A$
is a cross sectional area associated with each lattice bond.

\subsection{DC current}\label{subsec:stat-curr-ring}

In a static (time-independent) driving field $u$ the system reaches a
stationary state for long times, where the occupation probabilities
become constant, $p_i=p_i^{\rm st}$, and all local currents in
Eq.~(\ref{eq:local-j-ring}) are equal, $j_{i,i+1}=j_{\rm dc}$. Setting
$\kappa_i=1/\Gamma_{i+1}^-$ this leads to the recursion relation
\begin{equation}
p_{i+1}^{\rm st}=\eta_i p_i^{\rm st}- \kappa_i j_{\rm dc},
\label{eq:recursion-ring}
\end{equation}
with solution
\begin{equation}
p_i^{\rm st}= j_{\rm dc}\,\frac{\sum_{k=1}^M
\kappa_{i-k}\prod_{l=1}^{k-1}\eta_{i-l}}
{\prod_{k=1}^{M}\eta_{i-k}-1}\,.
\label{eq:pinf-ring}
\end{equation}
The current $j_{\rm dc}$ follows from the normalization,
\begin{equation}
\frac{1}{j_{\rm dc}}= \sum_{i=1}^N\frac{\sum_{k=1}^M
\kappa_{i-k}\prod_{l=1}^{k-1}\eta_{i-l}}
{\prod_{k=1}^{M}\eta_{i-k}-1}\,,
\label{eq:j-general-ring}
\end{equation}
which in turn fixes the occupation probabilities $p_i^{\rm st}$.
Equations~(\ref{eq:pinf-ring},\ref{eq:j-general-ring}) hold true for
arbitrary set of rates (as long as they do not exclude the formation
of a unique stationary state).

For detailed balanced rates these expressions can be simplified. With
condition (\ref{eq:detailed-balance}) we have
\begin{equation}
\prod_{l=1}^k \eta_{i-l}=\exp(\epsilon_{i-k}-\epsilon_i+ku)
\label{eq:prods}
\end{equation}
so that Eq.~(\ref{eq:j-general-ring}) can be written in the form
\begin{equation}
\frac{1}{j_{\rm dc}}=\frac{e^{-u/2} }{e^{Mu}-1}\sum_{l=1}^M e^{lu}
\sum_{k=1}^M\frac{ \exp[(\epsilon_k+\epsilon_{k+1})/2]}
{\sqrt{\Gamma_k^+(u)\Gamma_{k+1}^-(u)}} \exp(-\epsilon_{k+l})\,,
\label{eq:jst-ring}
\end{equation}
where we explicitly indicated the dependence of the jump rates
$\Gamma_k^\pm=\Gamma_k^\pm(u)$ on the external bias $u$. For the
coupling $\propto\exp(\pm u/2)$ of the rates to the external field it
can be shown that this formula agrees with Eq.~(10) in
Ref.~\onlinecite{Kehr/etal:1997} (or with Eqs.~(8-11) in
Ref.~\onlinecite{Heuer/etal:2005}).

In the linear response limit $u\to0$, Eq.~(\ref{eq:jst-ring}) reduces
to the result \cite{Maass/etal:1999} $qJ_{\rm
dc}/A=qnMj/A=\sigma_1E_0$ with
\begin{equation}
\sigma_1=\frac{nq^2a^2}{k_{\rm
    B}T}\left(\frac{1}{M}\sum_{k=1}^M\frac{1}{p_k^{\rm
      eq}\Gamma_k^+}\right)^{-1}\,,
\label{eq:sigma-lin-ring}
\end{equation}
where $p_k^{\rm eq}\propto\exp(-\epsilon_k)$ is the equilibrium
distribution and $\Gamma_k^+$ are the rates in the absence of
external driving ($u=0$). This formula can be viewed as resulting
from conductances $\propto p_k^{\rm eq}\Gamma_k^+$ in serial order.

In systems with only barrier disorder (all sites have the same energy
$\epsilon_k=0$), Eq.~(\ref{eq:jst-ring}) reduces to
\begin{align}
\frac{1}{j_{\rm dc}}&= \frac{1}{2 \sinh(u/2)} \sum_{k=1}^M \frac{1}
{\sqrt{\Gamma_k^+(u)\Gamma_{k+1}^-(u)}} \, .
\end{align}
Because $\Gamma_{k}^+ (-u)=\Gamma_{k+1}^- (u)$ in this case, we have
$\Gamma_{k}^+ (-u) \Gamma_{k+1}^- (u)=\Gamma_{k+1}^- (u) \Gamma_{k}^+
(u)=\Gamma_{k}^+ (u) \Gamma_{k}^- (u)$, and it follows that
$j(-u)=-j(u)$ for each disorder configuration. This is at first sight
a surprising results, since one could consider an asymmetric spatial
arrangement of barriers, for example, $U_{i,i+1}=iU_0$ for
$i=1,\ldots,N$ and $U_0>0$. If a particle would be driven in the
direction of increasing $i$, it encountered increasing barriers until
a jump from the largest to the smallest barrier occurs (after passing
the barrier $U_{N,N+1}=U_{N,1}$ between sites $N$ and 1). When driving
the particle in the reverse direction the opposite behavior would
results, i.e.\ the particle encountered smaller and smaller barriers
until a jump from the smallest barrier to the largest occurs.

Moreover, as long as the barriers for the local transitions are taken
into account by a simple Boltzmann factor, i.e.\ $\Gamma_k^+(u)\propto
\exp(-U_{k,k+1})f_+(u)$ and $\Gamma_{k+1}^-(u)\propto
\exp(-U_{k,k+1})f_-(u)$ with functions $f_\pm(u)$ independent of $k$,
one obtains the same current-voltage curve $J(u)=nMj(u)$ as in an
ordered system up to a rescaling factor. In such ordered system,
$\Gamma_k^+\Gamma_{k+1}^-=\Gamma^+\Gamma^-$ is independent of $k$, and
one obtains an $M$ independent total current $J_{\rm dc}=nMj_{\rm
  dc}$,
\begin{align}
J_{\rm dc}(u)&=
2n \sinh\left(\frac{u}{2}\right) \sqrt{\Gamma^+(u)\Gamma^-(u)} \, .
\label{eq:j-ordered}
\end{align}

\begin{figure}[!ht]
\centering
\includegraphics[width=0.48\textwidth,clip=,]{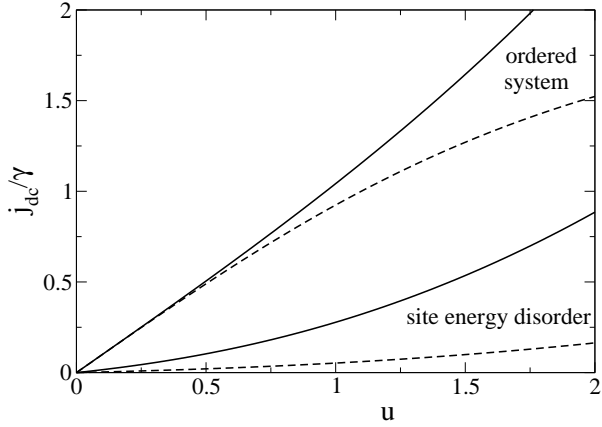}
\caption{\label{fig:1} Current $j_{\rm dc}(u)$ in the ring system
as a
  function of the bias $u$ for the exponential rates (solid line) and
  the Glauber rates (dashed line). Results are shown for an ordered
  system and a box distribution of site energies with
  $\Delta_\epsilon=6$. In the system with site energy disorder the
  mean current is shown, obtained after averaging $j_{\rm dc}$ from
  Eq.~(\ref{eq:j-general-ring}) over 100 realizations. In the case of
  pure barrier disorder the same curves as in the ordered system are
  obtained for each disorder realization up to a
  (realization-dependent) rescaling of the current (see text).}
\end{figure}

Figure~\ref{fig:1} shows the current $J_{\rm dc}(u)$ in the ordered
ring system (or in the ring systems with barrier disorder) for the
exponential rates (\ref{eq:exp-rates}) and the Glauber rates
(\ref{eq:glauber-rates}). For comparison we also show the average
current $J_{\rm dc}(u)$ in the case of the box distribution of site
energies with $\Delta_\epsilon=6$. The current was calculated
according to Eqs.~(\ref{eq:total_current},\ref{eq:jst-ring}) and
averaged over 10$^3$ different realizations of the site energy
disorder in rings with $M=10^3$ sites. The current-voltage curves in
the experimentally relevant regime $u\lesssim1$ tend to have a more
convex shape in the presence of site energy disorder. For the Glauber
rates the current is smaller and saturates for $u\to\pm\infty$.

%---------------------------------------------------------
%---------------------------------------------------------

In the ordered ring system, for the generic coupling
$\Gamma^{\pm}(u)=(\gamma/2)\exp(\pm u/2)$, one recovers from
Eq.~(\ref{eq:j-ordered}) the known result for the charge current
density \cite{Mott/Davis:1979}
\begin{align}
\frac{qJ_{\rm dc}}{A}=\frac{\gamma qn}{A}\sinh\left(\frac{u}{2}\right)=
\sigma_1 E + \sigma_3 E^3 + \mathcal{O}(E^5)
\end{align}
with $\sigma_1=(n/aA)\gamma q^2a^2/2k_{\rm B}T$ and
$\sigma_3=(n/aA)\gamma q^4a^4/48(k_{\rm B}T)^3$.
These results motivate to define an effective jump length by
\begin{equation}
a_{\rm eff}^2=\frac{24(k_{\rm B}T)^2}{q^2}\frac{\sigma_3}{\sigma_1}\,.
\label{eq:aeff}
\end{equation}

However, even in an ordered system it is possible that this effective
jump length does not yield a reasonable estimate of the true jump
length $a$. The reason is that, while the linear response quantity
$\sigma_1$ is universal (i.e.\ independent of the specific form of the
jump rates), this is not the case for the nonlinear conductivity
$\sigma_3$. For example, for the Glauber rates, we obtain $J=\gamma
n\tanh(u/2)$ from Eq.~(\ref{eq:j-ordered}) and accordingly a negative
$\sigma_3=-(n/aA)\gamma q^4a^4/24(k_{\rm B}T)^3$. If this would be
inserted in Eq.~(\ref{eq:aeff}), $a_{\rm eff}$ became imaginary.

In the general case, we can expand $\Gamma^+(u)$ in a Taylor series,
$\Gamma^+(u)=(\gamma/2)(1+\alpha_1u+\alpha_2u^2+\alpha_3u^3+\ldots)$.
{}From $\Gamma^-(u)=\Gamma^+(-u)=\exp(-u)\Gamma^+(u)$ it follows that
$\alpha_1=1/2$ independent of the specific form. With
Eq.~(\ref{eq:j-ordered}) we find
$\sigma_3/\sigma_1=(\alpha_2-1/12)q^2a^2/ (k_{\rm B}T)^2$, i.e.\
\begin{equation}
%a_{\rm eff}^2=\frac{1}{\left(\alpha_2-\frac{1}{12}\right)}
%\frac{(k_{\rm B}T)^2}{q^2}\frac{\sigma_3}{\sigma_1}\,.
a_{\rm eff}^2=24\left(\alpha_2-\frac{1}{12}\right)a^2\,.
\label{eq:aeff-2}
\end{equation}
We conclude that dependent on $\alpha_2$ (e.g., $\alpha_2=1/8$ for the
exponential rates, yielding $a_{\rm eff}=a$, and $\alpha_2=0$ for the
Glauber rates, yielding $a_{\rm eff}^2=-2a^2$) different $a_{\rm eff}$
can be obtained even in an ordered system.

\subsection{Thermodynamic limit and rectification}
\label{subsec:thermo-limit-ring}

In the thermodynamic limit $M\to\infty$ the sum over $k$ in
Eq.~(\ref{eq:jst-ring}) can be replaced by a disorder average
$\langle\ldots\rangle$ if the site energies and energy barrier do not
exhibit very broad distributions or long-range correlations, i.e.\ if
the the system is self-averaging. Accordingly we define
\begin{align}
a_l(u)&=\lim_{M\to\infty} \frac{1}{M}\sum_{k=1}^M\frac{
\exp[(\epsilon_k+\epsilon_{k+1})/2]} {\sqrt{\Gamma_k^+(u)\Gamma_{k+1}^-(u)}}
\exp(-\epsilon_{k+l})\nonumber\\
&=\left\{\begin{array}{l@{\hspace{1em}}l}
\displaystyle\left\langle
\frac{
\exp[(\epsilon_1-\epsilon_2)/2]}
{\sqrt{\Gamma^+(u;U_{12},\epsilon_1,\epsilon_2)
\Gamma^-(u;U_{12},\epsilon_1,\epsilon_2)}}
\right\rangle\,, & l=1\,,\\[4ex]
\displaystyle\left\langle
\frac{
\exp[(\epsilon_1+\epsilon_2)/2]\exp(-\epsilon_3)}
{\sqrt{\Gamma^+(u;U_{12},\epsilon_1,\epsilon_2)
\Gamma^-(u;U_{12},\epsilon_1,\epsilon_2)}}
\right\rangle\,, & l=2,\ldots\,M-1\,,\\[4ex]
\displaystyle\left\langle
\frac{
\exp[(\epsilon_2-\epsilon_1)/2]}
{\sqrt{\Gamma^+(u;U_{12},\epsilon_1,\epsilon_2)
\Gamma^-(u;U_{12},\epsilon_1,\epsilon_2)}}
\right\rangle\,, & l=M\,,
\end{array}\right.
\label{eq:al}
\end{align}
where we took into account the periodic boundary conditions and have
explicitly denoted the dependence of the jump rates
$\Gamma_k^{\pm}(u)=\Gamma^{\pm}(u;U_{k,k+1},\epsilon_k,\epsilon_{k+1})$
on the energies.

Keeping the number density $n$ fixed in the limit $M\to\infty$, we
then obtain from Eq.~(\ref{eq:jst-ring}) for the total current
\begin{align}
J_{\rm dc}(u)&=2n\sinh\left(\frac{u}{2}\right)
\frac{\displaystyle e^{|u|}}{\displaystyle
  [\theta(-u)a_1(u)+\theta(u)a_M(u)](e^{|u|}-1)+a_2(u)}\,,
\label{eq:jringlimit}
\end{align}
where $\theta(.)$ is the Heaviside step function [$\theta(x)=1$ for
  $x\ge0$ and zero else].
As discussed in the Introduction, Eq.~(\ref{eq:jringlimit})
should apply to typical experiments on thin film electrolytes.
There should be no notable dependence of the current (and the
nonlinear conductivities) on the film thickness, in agreement with
the experimental observations.

\begin{figure}[!ht]
\centering
\includegraphics[width=0.48\textwidth,clip=,]{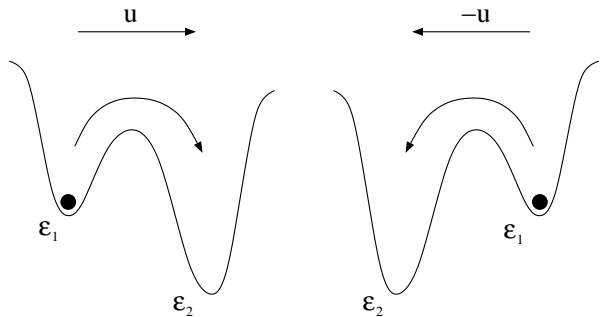}
\caption{\label{fig:2} Two mirror configurations with
  interchanged site energies $\epsilon_1$ and $\epsilon_2$ (and same
  $\epsilon_3$, not shown), as appearing with equal statistical weight
  in the averages in Eq.~(\ref{eq:al}). The jump rate
  $\Gamma^+(u,U_{12},\epsilon_1,\epsilon_2)$ in the left
  configuration is equal to the jump rate
  $\Gamma^-(-u,U_{12},\epsilon_2,\epsilon_1)$ in the right
  configuration after reversal of the bias $u$.}
\end{figure}

We can further show that the current from Eq.~(\ref{eq:jringlimit}) is
anti-symmetric with respect to the bias $u$. To this end we have to
analyze the symmetry properties of the $a_l(u)$. Note that in the
averages in Eq.~(\ref{eq:al}) there occur configurations with two or
three sites only, having mutually independent random site energies
$\epsilon_1,\ldots,\epsilon_3$. As illustrated in Fig.~\ref{fig:2}, to
each realization of the two energies $\epsilon_1$ and $\epsilon_2$
there exists a ``mirror configuration'' with interchanged site
energies $\epsilon_1$ and $\epsilon_2$, and the same value of
$\epsilon_3$. Since these mirror configurations occur with equal
statistical weight and exhibit the symmetry property
$\Gamma^+(u;U_{12},\epsilon_1,\epsilon_2)=
\Gamma^-(-u;U_{12},\epsilon_2,\epsilon_1)$, we can use
$\Gamma^+(u;U_{12},\epsilon_1,\epsilon_2)
\Gamma^-(u;U_{12},\epsilon_1,\epsilon_2)
=\Gamma^+(-u;U_{12},\epsilon_2,\epsilon_1)
\Gamma^-(-u;U_{12},\epsilon_2,\epsilon_1)$ in the averages of
Eq.~(\ref{eq:al}). This implies $a_1(-u)=a_N(u)$ and $a_2(-u)=a_2(u)$,
leading to $J(-u)=-J(u)$. Let us note that this does not imply that
the expansion of $J(u)$ contains odd powers of $u$ only. Terms
$\propto |u|^{2n+1}u$, $n=0,1,\ldots$, can occur according to
Eq.~(\ref{eq:jringlimit}) (see also the discussion in
Ref.~\onlinecite{Roling/etal:2008} for the consequences of these
non-analytic terms with respect to the analysis of experiments).

%-------------------------------------------------------------
%-------------------------------------------------------------

\begin{figure}[!ht]
\centering
\includegraphics[width=0.48\textwidth,clip=,]{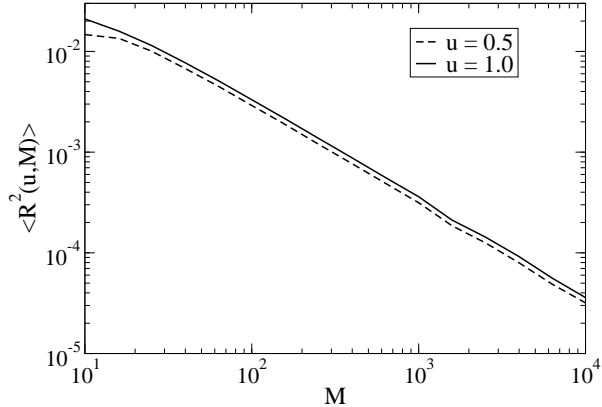}
\caption{\label{fig:3} Variance $\langle R^2(u,M)\rangle$ of the
  distribution of the rectification parameter $R(u,M)$ for the ring
  system in dependence of the system size $M$ at two fixed values of
  the bias $u$. The $R(u,M)$ were calculated from
  Eqs.~(\ref{eq:jst-ring},\ref{eq:rpara}) for a box distribution of
  site energies with $\Delta_\epsilon=6$ and disorder averages were
  performed over $10^3-10^5$ realizations.}
\end{figure}

In view of the antisymmetric current in the thermodynamic limit, we
expect, due to self-averaging, rectification effects for one system to
become smaller with increasing system size. To check this expectation,
we define the rectification parameter
\begin{align}
R(u,M)&=\frac{J_{\rm dc}(u)+J_{\rm dc}(-u)}{J_{\rm dc}(u)-J_{\rm dc}(-u)}
\label{eq:rpara}
\end{align}
for each disorder configuration in a ring with $M$ sites with $J_{\rm
  dc}(u)=nMj$ and $j_{\rm dc}$ from Eq.~(\ref{eq:j-general-ring}). The
distribution of this rectification parameter is, on symmetry reasons,
an even function of $u$, hence $\langle R(u,M)\rangle=0$. In the case
of self-averaging, the variance $\langle R(u,M)^2\rangle$ should
decrease as $\sim 1/M$ for $M\to\infty$. As shown in Fig.~\ref{fig:3},
this behavior is nicely confirmed by taking disorder averages of
$R^2(u,M)$.

%------------------------------------------------------------------------
%------------------------------------------------------------------------

\section{Conduction in open channels}\label{sec:channel}

So far we have considered ring systems with periodic boundary
conditions. In many situations the coupling of the system to particle
reservoirs is of importance, as in molecular wires, ion channels
through membranes, and thin-film electrolytes in contact with
non-blocking electrodes. In these systems details of the contact with
the reservoir can play a decisive role for the transport behavior, so
that a specific treatment is needed for the particular system under
consideration.

On the other hand, if one is interested in generic features of the
particle transport, one can adopt a coarse-grained description, where
only a few external parameters enter, as, for example, the
thermodynamic driving force of a reservoir to bring the system into
equilibrium with itself. Based on such coarse-grained description we
will in the following characterize a reservoir by its chemical
potential (amounting to a ``site energy level'' relative to the site
energies of the system), and an energy barrier for exchanging
particles between the system and the reservoir.

To be specific, we consider a one-dimensional channel consisting of
$M$ sites, which is coupled to sites $k=0$ and $k=M+1$, belonging to
two reservoirs with chemical potentials $\mu_{\rm L}^0=\epsilon_0$ and
$\mu_{\rm R}^0=\epsilon_{M+1}$, respectively. Particles are injected
or ejected from the two reservoir sites with rates that fulfill the
condition of detailed balance with respect to the grand-canonical
ensembles associated with $\mu_{\rm L}^0$ and $\mu_{\rm R}^0$. As for
the ring system, the site energies $\epsilon_k$ and the barrier
energies $U_{k,k+1}$, $k=0,\ldots,M$, determine the jump rates in the
absence of the external bias $u$, see Sec.~\ref{sec:transrates}
($U_{0,1}$ and $U_{M,M+1}$ specify the energy barriers for exchange of
particles with the left and right reservoir, respectively). In the
presence of a spatially uniform bias $u$, the potential drop along the
channel leads to the site energies
\begin{align}
E_k=\epsilon_k -ku\,,
\end{align}
and the electrochemical potentials
\begin{align}
\mu_{\rm L}=E_0=\mu_{\rm L}^0\qquad\mbox{and}\qquad
\mu_{\rm R}=E_{M+1}= \mu_{\rm R}^0-(M+1)u\,,
\end{align}
if we locate the point of zero external potential at the left end of
the channel. Note that for $k=0$ and $k=M$,
Eqs.~(\ref{eq:detailed-balance},\ref{eq:exp-rates},\ref{eq:glauber-rates})
define the jump rates for entering and leaving the system, in
agreement with detailed balance with respect to the grand-canonical
ensembles associated with $\mu_{\rm L}^0=\epsilon_0$ and $\mu_{\rm
  R}^0=\epsilon_{M+1}$.

In the open channel the particle number is a random variable and it is
not possible to consider a single-particle approach from the
beginning. The rate equations for the local concentrations
$p_i=\langle n_i\rangle$ follow from a Fermi lattice gas model, where
the occupation numbers $n_i$ at each site can have only two values
$n_i=0$ (vacant site) or $n_i=1$ (occupied site), and the set
$\{n_i\}$ specifies the microstate in the channel. The average
$\langle\ldots\rangle$ has to be taken with respect to the probability
distribution of the microstates at time $t$, whose time evolution
follows a master equation. Based on the master equation the derivation
of the currents $j_{i,i+1}$ in the equations of motions
(\ref{eq:rate-eq-ring}) is straightforward (for a systematic approach,
including also models with particle-particle interactions going beyond
site exclusion, see Ref.~\onlinecite{Gouyet/etal:2003}). The result is
\begin{align}
j_{i,i+1}&=
\Gamma^+_i\langle n_i(1-n_{i+1})\rangle-
\Gamma^-_{i+1}\langle n_{i+1}(1-n_i)\rangle\,,
\qquad i=1,\ldots,{M-1}\,.
\label{eq:bulkj}
\end{align}
For the boundary currents specifying the exchange of particles with
the reservoirs one obtains
\begin{subequations}
\label{eq:boundaryj}
\begin{align}
j_{0,1}
&=\Gamma_0^{+}(1-p_1)-\Gamma_1^{-}\,p_1\,,\label{eq:boundaryja}\\
j_{M,M+1}
&=\Gamma_M^{+}\,p_M-\Gamma_{M+1}^{-}(1-p_M)\,.
\label{eq:boundaryjb}
\end{align}
\end{subequations}

In a mean-field approximation, $\langle n_in_{i+1}\rangle\simeq\langle
n_i\rangle\langle n_{i+1}\rangle=p_ip_{i+1}$, the currents in
Eq.~(\ref{eq:bulkj}) can be expressed as
\begin{align}
j_{i,i+1}&=
\Gamma^+_ip_i(1-p_{i+1})-
\Gamma^-_{i+1}p_{i+1}(1-p_i)\,,
\qquad i=1,\ldots,{M-1}\,.
\label{eq:bulkj-mf}
\end{align}
In contrast to the ring system, the $p_k$ are no longer normalized,
but the mean number density $\bar p$ of particles is, for fixed energy
disorder, controlled by the electrochemical potentials $\mu_{\rm L}$
and $\mu_{\rm R}$. Accordingly, the currents $j_{k,k+1}$ in
Eqs.~(\ref{eq:bulkj-mf},\ref{eq:boundaryj}) are particle currents
(rather than probability currents) along the bonds between sites $k$
and $k+1$.

The nonlinear dependence on the $p_i$ leads, for non-vanishing bias
$u>0$, to interesting phase transitions of the mean particle
concentration with respect to variations of $\mu_{\rm L}$ and
$\mu_{\rm R}$, even in systems without energetic
disorder.\cite{comm:tasep} Based on exact solutions of the nonlinear
mean-field rate equations, one can show that these phase diagrams are
correctly predicted by the mean-field
approximation.\cite{Derrida:1998} The fact that phase transitions can
occur also in the dilute limit is sometimes disregarded. For example,
it has not been considered in treatments of incoherent hopping
transport of electrons along DNA molecules.

A thorough study of the nonlinear Eq.~(\ref{eq:bulkj-mf}) in the
presence of energetic disorder goes beyond the scope of this work. In
the special case of pure barrier disorder (all $\epsilon_i=0$) and a
current driven solely by a chemical potential difference
$\Delta\mu_0=\mu_{\rm L}^0-\mu_{\rm R}^0$ (bulk bias $u=0$), one has
$\Gamma_i^+=\Gamma_{i+1}^-$, and the nonlinear terms $\propto
p_ip_{i+1}$ in Eq.~(\ref{eq:bulkj-mf}) cancel. Accordingly, an
analytical solution of Eqs.~(\ref{eq:boundaryj},\ref{eq:bulkj-mf}) can
be obtained for the stationary state following the procedure discussed
in the following Sec.~\ref{subsec:stat-curr-channel}. The result for
the corresponding dc-current reads
\begin{align}
\label{eq:jdc-channel-symm}
J_{\rm dc}&=\frac{\Gamma^+_0\left(\Gamma^+_M+\Gamma^-_{M+1}\right)-\Gamma^-_{M+1}\left(\Gamma^+_0+\Gamma^-_1\right)}
{\left(\Gamma^+_0+\Gamma^-_1\right)+\left(\Gamma^+_M+\Gamma^-_{M+1}\right)+
\left(\Gamma^+_0+\Gamma^-_1\right)\left(\Gamma^+_M+\Gamma^-_{M+1}\right)\sum_{l=1}^{M-1}\frac{\displaystyle
  1}{\displaystyle\Gamma_l^+}}\,.
\end{align}
Note that due to the physical meaning of the $j_{k,k+1}$ discussed
above, the (total) current $J_{\rm dc}$
appears in Eq.~(\ref{eq:jdc-channel-symm}).

In the further treatment we will focus on situations where the
consideration of the one-dimensional geometry is an approximation for
a preferred bias direction of a higher-dimensional system, i.e.\ the
$p_i$ in Eqs.~(\ref{eq:boundaryj},\ref{eq:bulkj-mf}) are mean
concentrations (per site) that represent averages over a larger number
of sites belonging to lines or planes perpendicular to the current
direction. In this case we can, without worrying about the
boundary-induced phase transitions in one-dimensional geometries,
consider the dilute limit of Eq.~(\ref{eq:bulkj-mf}) with
$1-p_i\simeq1$,
\begin{align}
j_{i,i+1}&=
\Gamma^+_ip_i-
\Gamma^-_{i+1}p_{i+1}\,,
\qquad i=1,\ldots,{M-1}\,.
\label{eq:bulkj-mf-dilute}
\end{align}
The rate equations for the occupation probabilities $p_k(t)$ now have
the same form as in Eq.~(\ref{eq:rate-eq-ring}) for the
single-particle transport on the ring, but we have to take into
account the boundary currents according to Eq.~(\ref{eq:boundaryj}).
Moreover, one should keep in mind that the $p_i$, according to the
derivation of Eq.~(\ref{eq:bulkj-mf}), should be much smaller than
one. \cite{comm:dilute-limitation}

In total five external parameters control the transport behavior in
our model for the open channel: The chemical potentials $\mu_{\rm
  L}^0$ and $\mu_{\rm R}^0$, the energy barriers $U_{0,1}$ and
$U_{M,M+1}$ for particle exchange of the system with the reservoirs,
and the bias $u$. In the following, we will in most cases consider the
$\mu_{\rm L}^0$, $\mu_{\rm R}^0$, $U_{0,1}$, $U_{M,M+1}$ to be given
and discuss the transport behavior with respect to the bias $u$.

\subsection{DC current}\label{subsec:stat-curr-channel}

To calculate the stationary current under a static bias we iterate
Eq.~(\ref{eq:recursion-ring}) to obtain
\begin{align}
p_k^{\rm st} &= A_{k-1} p_1^{\rm st} -
J_{\rm dc}B_{k-1}\,,
\label{eq:p-recur}
\end{align}
with
\begin{align}
A_k &= \prod_{l=1}^{k}
\eta_l
%\frac{\Gamma_l^+ (u)}{\Gamma_{l+1}^- (u)}
=\exp[(E_1 - E_{k+1})]\,,\label{eq:coeffA} \\
B_k &= \sum_{m=1}^{k}
%\frac{1}{\Gamma_{m+1}^- (u)}
\kappa_m
\prod_{l=m+1}^{k}
\eta_l
%\frac{\Gamma_l^+(u)}{\Gamma_{l+1}^ -(u)}
= \sum_{m=1}^k \kappa_m \exp(E_{m+1}-E_{k+1})\,,
%\frac{\exp(E_{m+1}-E_{k+1})}{\Gamma_{m+1}^-}\,,
\label{eq:coeffB}
\end{align}
where the expression containing the products hold true in general,
while the second expressions are valid for detailed
balanced rates.

Using Eqs.~(\ref{eq:boundaryj}) together with Eq.~(\ref{eq:p-recur})
for $k=M$ one obtains a closed equation for $J_{\rm dc}$ with solution
\begin{subequations}
\label{eq:j_dc_channel}
\begin{align}
\label{eq:j_dc_channel-a}
J_{\rm dc}&=
\frac{1-\exp[-(\mu_L^0-\mu_R^0)-(M+1)u]}
{\displaystyle\frac{1}{\Gamma_0^+}+
\frac{\exp[-\epsilon_M-Mu-\mu_L^0])}{\Gamma_M^+}+
\sum_{k=1}^{M-1}\frac{\exp(\epsilon_k-\mu_{\rm L}^0-ku)}{\Gamma^+_k}}\\
\label{eq:j_dc_channel-b}
&=\frac{1-\exp{(-\Delta\mu)}}
{\displaystyle\sum_{k=0}^M\frac{\exp(E_k-\mu_{\rm L})}{\Gamma^+_k}}\,,
\end{align}
\end{subequations}
where $\Delta\mu=\Delta\mu_0-(M+1)u$. This result in turn fixes the
local concentrations $p_k^{\rm st}$ in Eq.~(\ref{eq:p-recur}) via
Eqs.~(\ref{eq:coeffA},\ref{eq:coeffB}) and $p_1^{\rm st}$ from
Eq.~(\ref{eq:boundaryja}) with $j_{0,1}=J_{\rm dc}$.

Equation~(\ref{eq:j_dc_channel-b}) may be interpreted in a similar way
as the linear response in the ring system, cf.\
Eq.~(\ref{eq:sigma-lin-ring}): The current follows from a driving
force $1-\exp(-\Delta\mu)$ and a total ``conductance'' given by
elementary ``conductances'' $\exp[-(E_k-\mu_{\rm L})]\Gamma_k^+$ in
serial order. Equation~(\ref{eq:j_dc_channel-b}) is, however, not a
linear response formula, but describes the full nonlinear response to
the bulk driving force $u$ and the boundary driving force
$\Delta\mu_0$. Note that these driving forces do not enter
Eq.~(\ref{eq:j_dc_channel-b}) in the single combination
$\Delta\mu=\Delta\mu_0-(M+1)u$, since $\Gamma_0^+$ and $\Gamma_M^+$
are controlled independently by $\mu_{\rm L}^0$ and $\mu_{\rm R}^0$,
respectively.

Due to the factors $\propto\exp(-ku)$ in
Eq.~(\ref{eq:j_dc_channel-a}), only jump rates $\Gamma_k^+$
($\Gamma_{M-k}^+$), $k=1,2,\ldots$, from sites close to the left
(right) boundary give a significant contribution for positive
(negative) bias $u$. This means that for $u\ne0$, $J_{\rm dc}$ is
governed by jump rates belonging to sites in a region of size $\propto
1/u$ close to either boundary. As a consequence, already pure barrier
disorder (with all $\epsilon_k=0$) leads to rectification effects in
the open channel, in marked contrast to the behavior in the ring
system.

\begin{figure}[!ht]
\centering
\includegraphics[width=0.48\textwidth,clip=,]{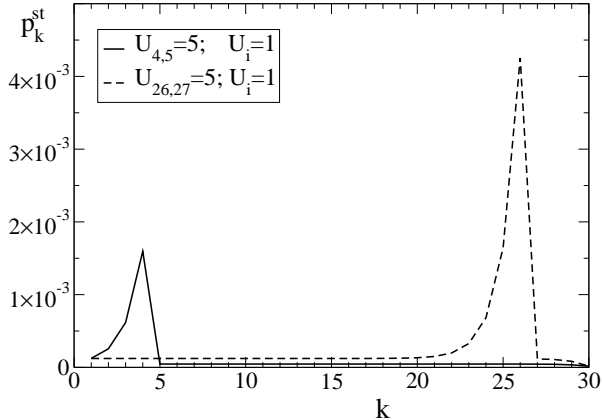}
\caption{\label{fig:4} Stationary density profiles in an open
channel
  with $M=30$ sites, a constant bias $u=1$, and a single large barrier
  $U_{l,l+1}=5$ close to the left ($l=4$, solid line) and close to the
  right boundary ($l=26$, dashed line); the other barriers are set to
  one, $U_{k,k+1}=1$ for $k\ne l$ (including the boundary barriers for
  exchange of particles with the reservoirs with
  $\mu^0_R=\mu^0_L=-10$), and all $\epsilon_k=0$, $k=0,\ldots,M+1$. As
  a consequence of the density profile, the current
  $J=1.8\cdot10^{-5}$ for the large barrier at site $l=4$ is smaller
  than the the current $J=4.7\cdot 10^{-5}$ for the larger barrier at
  site $l=26$. In the latter case $J$ has practically the same value
  as in the corresponding ordered system (all $U_{k,k+1}=1$).}
\end{figure}

It may be surprising at first sight that the dominant contribution to
the current comes from regions close to either boundary (for similar
phenomena expected in connection with electron transport though
molecular bridges, see Ref.~\onlinecite{Datta:2005}). The effect can
be understood when considering, without generality, $u>0$, and a
single large barrier $U_{l,l+1}>U_0$ in an otherwise ordered system
with smaller barriers $U_{k,k+1}=U_0$ for $k\ne l$ (and all
$\epsilon_k=0$). Let us first look at the density profile in the
region of sites left [$k\le l$] and right [$k\ge (l+1)$] of the large
barrier. For the current $j_{l,l+1}>0$ across the large barrier
$U_{l,l+1}$ to equal all other currents $j_{k,k+1}$, the
concentrations $p_k$ in the right region have to be much smaller than
$p_j$, while the local concentrations in the left region must decrease
smoothly with increasing distance from the large barrier (smaller
$k$). Hence the density profile in the stationary state has a maximum
at site $j$ with a smooth decay to the left and a sharp fall to the
right of the large barrier $U_{j,j+1}$. This is demonstrated in
Fig.~\ref{fig:4}, where we show the solution $p_k^{\rm st}$ for a
large barrier close to the left boundary (solid line) and close to the
right boundary (dashed line). As a consequence, when the large barrier
is closer to the left boundary, the density at the boundary site $k=1$
becomes larger, leading to a smaller current $J_{\rm dc}=j_{0,1}$.
More generally speaking, we can say that for $u>0$ ($u<0$) the energy
landscape close to the left (right) boundary controls the density at
the boundary site $k=1$ ($k=M$) and thus the current $J_{\rm
  dc}=j_{0,1}$ ($J_{\rm dc}=j_{M,M+1}$). We note that the dominance of
the boundary regions will no longer apply when considering the
transport with site exclusion in strictly one-dimensional topologies
(ASEPs or TASEPs).

%----------------------------------------------------------------
%-----------------------------------------------------------------

\begin{figure}[!ht]
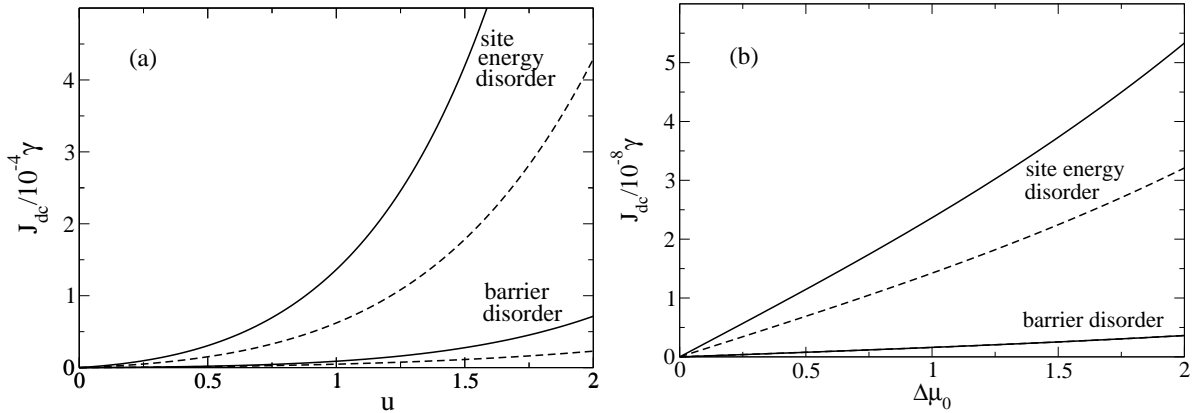

%\vspace*{-2ex}
\includegraphics[width=0.48\textwidth,clip=,]{fig5a}
\includegraphics[width=0.47\textwidth,clip=,]{fig5b}
\caption{\label{fig:5} Current $J_{\rm dc}$ (a) as a function of
  the bias $u$ at fixed $\mu_{\rm L}^0=\mu_{\rm R}^0=-10$, and (b) as
  a function of the chemical potential difference $\Delta\mu^0$ for
  vanishing bias $u=0$. Averages have been performed over 100
  realizations of the disorder, for a box distribution of energy
  barriers with $\Delta_U=5$, and a box distribution of site energies
  with $\Delta_\epsilon=6$. Solid lines refer to the exponential rates
  and dashed lines to the Glauber rates. In the case of barrier disorder
  and $u=0$, the exponential and Glauber jump rates are the same, and hence
  the corresponding currents agree in part (b).}
\end{figure}

To illustrate typical behaviors of the current, we calculate $J_{\rm
  dc}$ as a function of the driving forces for only barrier disorder
(all $\epsilon_k=0$) and for only site energy disorder (all
$U_{k,k+1}=0$), using the box distributions introduced in
Sec.~\ref{sec:transrates}. Figures~\ref{fig:5} show results for the
disorder averaged current (a) as a function of $u$ for $\mu_{\rm
  L}^0=\mu_{\rm R}^0=-10$, and (b) as a function of $\Delta\mu_0$ for
$\bar\mu^0=-10$ ($\mu_{\rm L,R}^0=-10\pm\Delta\mu_0$) and $u=0$.
Similar as in the ring system, the current-voltage curves in
Fig~\ref{fig:5}a have a more convex shape in the presence of site
energy disorder for small $u$. One may ask if the current $j_{\rm
  dc}^{\rm ring}$ in the ring system [Eq.~(\ref{eq:jst-ring})] and the
current $J_{\rm dc}^{\rm ch}$ in the open channel
[Eq.~(\ref{eq:j_dc_channel})] can be connected by simply taking
account the mean number $\bar N=\sum_{k=1}^M p_k^{\rm st}$ of
particles in the channel, i.e.\ if $J_{\rm dc}^{\rm ch}=\bar N J_{\rm
  dc}^{\rm ring}$. However, the fact that regions close to either
boundary govern the value of $J_{\rm dc}^{\rm ch}$, already shows that
such mapping cannot be correct. Indeed, based on the analytical
results (\ref{eq:jst-ring},\ref{eq:j_dc_channel}) obtained for the
ring system and open channel, one can show that such a relation does
not hold true. Numerical solutions also show that the relation does
not provide a reasonable approximation (see also the discussion in
Sec.~\ref{sec:time-dependent-response}).

%--------------------------------------------------------------
%--------------------------------------------------------------

\subsection{Thermodynamic limit and rectification}
\label{subsec:thermo-limit-channel}

The dominance of the boundary regions implies that the thermodynamic
limit has to taken is such a way that for $u>0$ the left boundary has
to be fixed and the right boundary goes to infinity, while for $u<0$
one should consider the reversed situation (fixed right boundary and
left boundary going to infinity). We focus on the case $u>0$ here
(with obvious analogous treatment for the case $u<0$). For
$M\to\infty$, Eq.~(\ref{eq:j_dc_channel-a}) then becomes
\begin{align}
J_{\rm dc}
&=\frac{1}
{\displaystyle\sum_{k=0}^\infty\frac{\exp(\epsilon_k-ku-\mu_{\rm L})}
{\Gamma^+_k}}
=\frac{1}
{\displaystyle\sum_{k=0}^\infty\frac{\exp(E_k-\mu_{\rm L})}{\Gamma^+_k}}\,.
\end{align}
One can proof that for point-symmetric energy landscapes
($\epsilon_k=\epsilon_{M+1-k}$, $U_{k,k+1}=U_{M+1-k,M-k}$) the current
is antisymmetric with respect to a reversal of the driving forces,
i.e.\ $J_{\rm dc}(-u,\mu_{\rm L}\to\mu_{\rm R},\mu_{\rm R}\to\mu_{\rm
  L})=-J(u,\mu_{\rm L},\mu_{\rm R})$ (the reference point of zero external
potential has to be shifted from the left to right boundary also).

\begin{figure}[!ht]
\centering
\includegraphics[width=0.48\textwidth,clip=,]{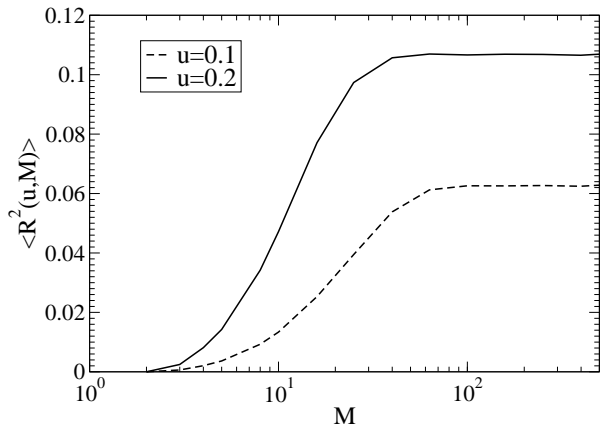}
\caption{\label{fig:6} Variance $\langle R^2(u,M)\rangle$ of the
  distribution of the rectification parameter $R(u,M)$ for the open
  channel in dependence of the system size $M$ at two fixed values of
  the bias $u$. The $R(u,M)$ were calculated from
  Eqs.~(\ref{eq:j_dc_channel},\ref{eq:rpara}) for a box distribution
  of site energies with $\Delta_\epsilon=6$ and disorder averages were
  performed over $10^3-10^5$ realizations.}
\end{figure}

Moreover, as mentioned above, rectification effects occur already for
pure barrier disorder and do not become smaller for increasing $M$.
Accordingly, the width of the distribution of the rectification
parameter defined in Eq.~(\ref{eq:rpara}) should saturate to a finite
value for $M\to\infty$. This is confirmed in Fig.~\ref{fig:6}, where
for pure energy disorder, $\langle R^2(u,M)\rangle$ is shown as a
function of $M$ for two fixed values of $u$ and $\mu_{\rm
  L}^0=\mu_{\rm R}^0=-10$. It would be interesting to check this
theoretical prediction in experiments, e.g.\ in thin film electrolytes
contacted to non-blocking electrodes. Systematic measurements in
dependence of the system size (film thickness) would allow one to
distinguish between a possible finite size effect and the effects
induced by the open boundaries.

%-----------------------------------------------------------------
%-----------------------------------------------------------------

\section{Time-dependent nonlinear response}
\label{sec:time-dependent-response}

In this section we discuss the time-dependent nonlinear response to a
sinusoidal electric field $E(t)=E_0\sin(\omega t)$ with large
amplitude $E_0$, corresponding to a bias $u(t)=u_0\sin(\omega t)$ with
amplitude $u_0=qE_0a/k_{\rm B}T\gtrsim1$. To this end we solve the
rate equations (\ref{eq:rate-eq-ring}) supplemented by periodic
boundary conditions for the ring and Eqs.~(\ref{eq:boundaryj}) for the
open channel. After a transient time interval the stationary regime is
reached, where we determine the total current $J_{\rm st}(t)$ averaged
over many periods. Fourier decomposition of this stationary current
yields the complex first order and higher harmonics $\hat
J_n(\omega)=J_n'(\omega)+iJ_n^{\prime\prime}(\omega)$, $n=1,2,\ldots$

In the high and low frequency limits the current $J_{\rm st}(t)$ (and
hence the harmonics $\hat J_n(\omega)$) can be calculated
analytically. For $\omega\to\infty$ and barrier disorder, the mean
local densities $p_i(t)$ in the stationary state become independent of
position and time, i.e.\ $p_i(t)=p$,\cite{comm:hf-limit} and one can
show that for each realization
\begin{align}
J_{\rm st}(t)&=\frac{\gamma
p}{2M}\left[\sum_{k=1}^M\exp(-U_{k,k+1})\right][f_+\left(u(t)\right)-f_-\left(u(t)\right)],
\label{eq:jhf}
\end{align}
where $f_\pm(u)$ are the factors modifying the transitions due to the
external driving [see discussion before Eq.~(\ref{eq:j-ordered})].
Upon averaging over the disorder (or due to self-averaging),
$\sum_{k=1}^M\exp(-U_{k,k+1})/M$ can be replaced by the ensemble
average $\langle\exp(-U_{1,2}\rangle$.

For $\omega\to0$, one can take the quasistatic limit,
\begin{align}
J_{\rm st}(t)&=J_{\rm dc}\left(u(t)\right)
\label{eq:jquasi}
\end{align}
with $J_{\rm dc}(.)$ from Eq.~(\ref{eq:jst-ring}) for the ring system
and Eq.~(\ref{eq:j_dc_channel}) for the open channel. For exploring
the intermediate frequency behavior we have to rely on our numerical
solution of the underlying rate equations.

In the following we will concentrate on barrier disorder, implying
that harmonics of even order vanish in the ring due to the absence of
rectification (see the discussion in
Sec.~\ref{subsec:thermo-limit-ring}). In the open channel, by
contrast, rectification effect are present and the harmonics of even
order are nonzero. However, these harmonics of even order are much
smaller than the harmonics of odd order, and therefore will not be
shown here. For the discussion of the harmonics of odd order we focus
on the real parts $J_n'(\omega)$.

\begin{figure}[!ht]
\centering
\includegraphics[width=0.48\textwidth,clip=,]{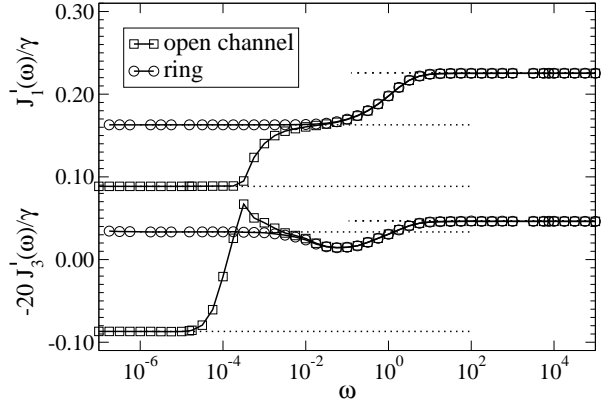}
\caption{\label{fig:7} First order and third order harmonics of
the
  current in the ring and open channel for $\Delta_U=2$,
  $M=2000$, and
  the exponential jump rates ($\mu_{\rm L}=\mu_{\rm R}=-1$ and
  $U_{0,1}=U_{M,M+1}=2$ for the
  open channel). The results have been averaged over the same sets of
  5 realizations of barrier disorder, and the single-particle results
  for the ring are matched to the mean particle number in the open
  channel. The dotted lines mark the limiting behavior for high and
  low frequency (see text).}
\end{figure}

Figure~\ref{fig:7} shows the harmonics $J_1'(\omega)$ and
$J_3'(\omega)$ in the case of the exponential jump rates for the
barrier disorder with $\Delta U=2$ and bias amplitude $u_0=1$ (for the
channel we have set $\mu_{\rm L}^0=\mu_{\rm R}^0=-1$ and boundary
barriers $U_{0,1}=U_{M,M+1}=2.2$. The results were averaged over 5
realization of the disorder. The circles mark the results for the ring
system and the squares for the open channel.

%----------------------------------------------------------------
%----------------------------------------------------------------

In the ring system, the first harmonics $J_1'(\omega)$ shows the
typical behavior known for a hopping system in the linear response
limit: In a high frequency regime, $J_1'(\omega)$ shows a plateau, and
then, upon lowering the frequency, it decreases monotonously within a
dispersive regime until approaching the low-frequency regime, where
$J_1'(\omega)$ again becomes independent of $\omega$. The third order
harmonics $J_3'(\omega)$ in the ring also shows a plateau at high and
low frequencies, and passes through a minimum in the dispersive
regime. The plateau values in the limits of high and low frequencies
follow from Eq.~(\ref{eq:jhf}) and Eq.~(\ref{eq:jquasi}),
respectively, and are marked by dotted lines in the figure. With
respect to the imaginary parts $J_1^{\prime\prime}(\omega)$ and
$J_3^{\prime\prime}(\omega)$, we found peaks appearing in the
dispersive regimes in Fig.~\ref{fig:7}.

In the open channel the harmonics follow those in the ring system for
higher frequencies. This can be understood from the fact that at
higher frequencies the dynamics in the interior of the channel is
dominant (``bulk behavior''). At lower frequencies, however, the
coupling to the reservoirs leads to significant changes in the mean
particle number. As a consequence, an additional dispersive
regime\cite{comm-additional} is seen at low frequencies, until the
limit corresponding to Eq.~(\ref{eq:jquasi}) is reached. Note in
particular that $J_3'(\omega)$ changes its sign when approaching the
low-frequency limit.

Let us finally note that we have obtained an analogous overall
behavior of the harmonics in the case of site energy disorder with the
notable difference that no change of sign in $J_3'(\omega)$ was
observed.

\section{Summary and Conclusions}

The problem of one-dimensional hopping transport has gained renewed
interest, in particular in connection with biophysical applications
and electron transport through molecular wires. We have discussed in
this work the situation for non-interacting particles with a focus on
disorder effects (or regular variations of site and barrier energies)
on the current response to an external bias. For both the periodic
ring system and the open channel analytical results were derived for
the stationary current in response to static external driving forces,
without making specific assumptions on the form of the jump rates.
Representative results were shown for spatially uncorrelated energy
landscapes, characterized by box distributions either in the barrier
or site energies.

It was further shown that in the ring system rectification effects
become smaller for increasing system size. In the thermodynamic limit
of infinite system size, the current $J_{\rm dc}(u)$ becomes
anti-symmetric with respect to the bias $u$ and its expansion in
powers of $u$ can exhibit non-analyticities of the form $|u|^{2n+1}u$,
$n=0,1,\ldots$ In the open channel rectification does not vanish in
the thermodynamic limit due to the fact that the current is dominated
by the variations of the energy landscape close to either system
boundary dependent on the bias direction. It would be interesting to
check this rectification effect in experiments, as, for example, in
measurement of ionic currents in electrolytes in contact with
non-blocking electrodes.

Numerical solutions of the underlying rate equations were obtained for
a sinusoidal external driving and results were presented for the first
and higher harmonics of the current. For intermediate and high
frequencies the harmonics in the open channel were shown to equal
those in the ring, if the particle concentration is adapted properly.
In the low-frequency regime the harmonics can be derived from the
quasistatic limit. This implies that the low-frequency limit is
different in the open channel from that in the ring. The origin of
this difference can be attributed to changes in the mean particle
number in the open system, which are not present in the ring model.

The results presented here provide a basis for further investigations
of interacting particles. As discussed in Sec.~\ref{sec:channel}, in
truly one-dimensional geometries already hard-core interactions can
change the general characteristics of the transport behavior due to
boundary induced phase transitions of the mean particle concentration.
Influences of disorder effects on these phase transitions have been
discussed in various works (see e.g.\
Refs.~\onlinecite{Tripathy/etal:1997,Harris/etal:2004,Evans/etal:2004}),
but a thorough general treatment for arbitrary disorder has not been
provided yet. Only a few studies have been performed for longer range
particle-particle interactions. An example is the treatment of
nearest-neighbor repulsions in TASEPs on the basis of specific rules
for the transition rates.\cite{Krug:1991,Hager/etal:2001} This can
give rise to more complex phase diagrams compared to the case of
hard-core interactions. A more complete exploration of the effects of
disorder and particle-particle interactions, as required to get a more
detailed description of real systems, still remains an open challenge.

\acknowledgments We thank W.~Dieterich for very valuable discussions.
Parts of this work were supported by the HI-CONDELEC EU STREP project
(NMP3-CT-2005-516975).


\begin{thebibliography}{99}
\bibitem{comm:hf-limit} This does no longer hold true in the case of
  energy disorder so that a more sophisticated treatment has to be
  performed in this case.
\bibitem{comm-additional} For stronger disorder, different system
  sizes and different chemical potentials, the onset frequency, where
  the behavior in the open channel starts to deviate from the behavior
  in the ring, can get shifted.
\bibitem{comm:dilute-limitation} The set of equations
  (\ref{eq:boundaryj},\ref{eq:bulkj-mf-dilute}) is, of course,
  intuitive and one may use it as a starting point of a
  phenomenological description without referring to its derivation
  from the fermionic lattice gas. In such phenomenological approach
  one would not necessarily require the $p_i$ to be much smaller than
  one. Indeed, we found that due to the linearity of the equations of
  motions, this restriction is not relevant with respect to the
  overall qualitative behavior of the system. However, application of
  these equations to situations with $p_i$'s not being much smaller
  than one should be considered with care.
\end{thebibliography}
\end{document}